\documentclass[runningheads]{llncs}

\usepackage[title]{appendix}
\usepackage{graphicx}
\usepackage{comment}
\usepackage[subrefformat=parens]{subcaption}
\begin{document}
\title{
``When and where do you want to hide?"\\-- Recommendation of location privacy preferences with local differential privacy}
\titlerunning{Recommendation of location privacy preferences with LDP}
%
\author{Maho ASADA \and
Masatoshi YOSHIKAWA \and
Yang CAO}
\authorrunning{M. Asada et al.}
%
\institute{Kyoto University, Kyoto, Japan\\
\email{asada@db.soc.i.kyoto-u.ac.jp}\\
\email{\{yoshikawa, yang\}@i.kyoto-u.ac.jp}}
\maketitle              
\begin{abstract}

In recent years, it has become easy to obtain location information quite precisely.
However, the acquisition of such information has risks such as individual identification and leakage of sensitive information, so it is necessary to protect the privacy of location information.
For this purpose, people should know their location privacy preferences, that is, whether or not he/she can release location information at each place and time.
However, it is not easy for each user to make such decisions and it is troublesome to set the privacy preference at each time.
Therefore, we propose a method to recommend location privacy preferences for decision making.
Comparing to existing method, our method can improve the accuracy of recommendation by using matrix factorization and preserve privacy strictly by local differential privacy, whereas the existing method does not achieve formal privacy guarantee.
In addition, we found the best granularity of a location privacy preference, that is, how to express the information in location privacy protection.
To evaluate and verify the utility of our method, we have integrated two existing datasets to create a rich information in term of user number.
From the results of the evaluation using this dataset, we confirmed that our method can predict location privacy preferences accurately and that it provides a suitable method to define the location privacy preference.
\end{abstract}
\section{Introduction}\label{sec-intro}

    In recent years, due to the popularization of smartphones and the development of GPS positioning equipment, location information for people has been able to be obtained quite precisely and easily.
    Such data can be utilized in various fields such as marketing and urban planning.
    In addition, there are many applications that do not function effectively without location information \cite{yokoyama2015real} \cite{kanza2015online} \cite{yokoyama2015real}.
    Because of such value, market maintenance to buy and sell it has started. \cite{databank} \cite{kanza2015online}.

    However, on the other hand, by publishing accurate location information, there are privacy risks associated with such as individuals being identified  \cite{de2013unique}.
    Due to such risks, privacy awareness regarding location information among people is very high \cite{chosa}.
    One of the most risky situations is when smartphones are used.
	This is because we are sending location information to them when using many applications \cite{fawaz2014location}.
	
    In order to prevent privacy risks under such circumstances, it is necessary to anonymize or obfuscate location information.
	One of the countermeasures is a primitive one: turn off location information transmission manually when using a smartphone.
	There is also a method of applying a location privacy protection technique.
	Various techniques are used for protection, including $k$-anonymity \cite{sweeney2002k} \cite{Huo2012YouCW}, differential privacy \cite{dwork2008differential} \cite{Jiang2013PublishingTW}, and encryption \cite{wasef2010rep}.
	Fawaz  \cite{fawaz2014location} proposed a system that applies these privacy protection technologies to smartphones.
    This system controls the accuracy of the location information sent to each application.
    The user needs to input how accurate he/she wants to send the respective location information to each application.
	
	In these countermeasures, to avoid a privacy risk by the disclosure of location information, the user has to decide location privacy preference for each location, that is, whether or not he/she publishes the location data at a certain place and time.
	However, we think that there is a problem in such a situation.
	What is the best privacy preference is unclear, and it may be different for each user.
	Therefore, individual users need to determine their location privacy preferences.
	However, most users find it difficult to determine these preferences themselves \cite{sadeh2009understanding}, and it is troublesome to set the privacy preference at each time.
	Of course, it is necessary to avoid sending all the position information to the third party such as application.
	
	Therefore, we need a system to recommend location privacy preferences for decision support when choosing a user's location information privacy preference and for the promotion of safe location information release.
	Recently, one such system was developed using the concept of item recommendation, which is used for online shopping.
	Item recommendation regards the combination of location and time as an item and whether or not to release location information as a rating of the item, and it predicts the rating of an unknown item.
	Zhang  \cite{zhao2014privacy} proposed a method of recommending by collaborative filtering.
	
    We focus on the problems of existing location privacy preference recommendation methods and propose a recommendation method to solve these problems.
	The contributions of this research are as follows.
	
	\begin{enumerate}
		\item {\bf Clarifying the definition of location privacy preference: }
		In location privacy preference, how to define time and place has not been clarified so far.
		For example, regarding time, ``2 pm, February 1st, 2019" can be represented  as ``a Friday afternoon of February", ``an afternoon", and so on.
		Since granularity affects the usefulness of a recommendation, it is necessary to confirm the best method of expression.
		Although many location privacy protection methods have been proposed in the literature, none of them addresses the problem of how to set location privacy preference.
        Therefore, we generate recommendation models using various granularities for time information and compare their usefulness.
        From these results, we find the best granularity that will produce trade-offs between the density of special data for the recommendation and the consideration of time.
		
		\item {\bf Applying matrix facrorization to location privacy preference recommendation: }
        It is very important to recommend accurately, since location privacy preferences are very sensitive.
		Collaborative filtering, which was used in the method by Zhang  \cite{zhao2014privacy}, experiences problems that when the number of users and products increase; accurate prediction cannot be achieved, and only the nature of either the user or product can be considered well.
		Therefore, we propose a method to improve by utilizing matrix factorization.
		As a result of experiments, we confirm that we can predict accurate evaluation values with a probability of 90\% for large amount of data.
		
		\item {\bf Recommendation with location differential privacy: }
		Matrix factorization is involved in privacy risk, because each user needs to send their data to the recommendation system.
		Location privacy preferences are very sensitive because they encompass location information of the users at a certain times and whether the information is sensitive for him/her.
		The recommendation system obtaining such data may be on an untrusted server through which an attacker may try to extract the user's data \cite{frey2015collaborative}.
		However, a location privacy preference recommendation that achieves highly accurate privacy protection has not been proposed so far.
		Actually, the method by Zhang  \cite{zhao2014privacy} did not preserve privacy in a strict mathematical sense.
		Therefore, we propose a recommendation method that realizes it with local differential privacy.
		In this technique, each user adds noise to their own data when transmitting data, and the technique enables sending data to any system.
		We confirm that our method maintain precision that is the same as that achieved a method without privacy protection.
		
		\item {\bf Generating a location information privacy preference dataset: }
		A challenge in experiments for testing the performance of our methods is that no appropriate location privacy preference dataset is available in literature.
		The only available real-world dataset of location privacy preference\cite{locShare} has few users.
		Such data is not suitable for the evaluation of the method using matrix factorization \cite{koren2009matrix}.
		In addition, we need bulk data in the evaluation because there are many users of recommendation in the real world.
		Therefore, we created an artificial dataset that combines such the location privacy preference dataset and a trajectory dataset with a large number of users.
	\end{enumerate}
	
	This paper is organized as follows:
	We describe the related works in Section \ref{sec-relatedwork} and the knowledge necessary for realizing our goal in Section \ref{preliminaries}.
	Then, we describe our method in Section \ref{sec-method} and evaluate and discuss about our method in Section \ref{sec-eval}.

\section{Related work}\label{sec-relatedwork}
	
	\subsection{Recommendation of location privacy preferences}
	
        In order to protect the privacy of the location information, it is necessary to know the location privacy preferences, but this task is difficult.
        The recommendation of location privacy preferences is useful in such conditions.
        There are two main types of recommendations: one that predicts preferences based on only one user's data \cite{fang2010privacy} \cite{sadeh2009understanding} and one that uses data of multiple users \cite{toch2014crowdsourcing} \cite{jin2013recommendations}.
        
        With the former type of recommendation, Sadeh \cite{sadeh2009understanding} proposed a recommendation method that used machine learning such as the random forest algorithm.
        However, in a system using only one user's data in prediction, when the user has few data about his/her privacy preference, and the system cannot perform accurately \cite{zhao2014privacy}.
        Such a problem is very likely to occur, since it is difficult for users to determine their own location privacy preferences themselves.
        
        A recommendation based on multiple users' data can solve such a problem.
        There is a method that applies item recommendations used in online shopping to the recommendation of location privacy preferences.
        This method considers the combination of location and time as an item in the recommendation and predicts the ratings of unknown items using collaborative filtering \cite{resnick1994grouplens} \cite{zhao2014privacy}.
        However, there are three disadvantages to this method: the accuracy decreases as the number of users and products increases, only the nature of either the user or product can be considered well, and there is a possibility of threatening the user's privacy.
        Therefore, we propose a location privacy preference recommendation system that improves the utility and preserves privacy.

	\subsection{Privacy preserving recommendations}\label{subsec-ppr}
	    Item recommendation has privacy risks \cite{calandrino2011you} \cite{lam2006you} \cite{mcsherry2009differentially}, 
		so the recommendation methods that aim to preserve privacy use differential privacy \cite{balu2016differentially} \cite{berlioz2015applying} \cite{liu2015fast} \cite{machanavajjhala2011personalized}.
		In these methods, the recommendation system adds noise after collecting users' ratings of items, and such a system is considered to be on a trusted server.
		However, there is no guarantee that all recommendation systems exist on a trusted server, and there are cases when privacy is threatened.
		
		To avoid such risks, local differential privacy is used, where the user adds noise to his/her data before sending the data to the recommendation system.
		The number of methods that apply local differential privacy to item recommendation is increasing \cite{shen2014privacy} \cite{shen2016epicrec} \cite{hua2015differentially}.
		The method by Shin \cite{Shin2018PrivacyEM} added both items and rating, improving the privacy preservation.
		
		However, there has never been a location privacy preference recommendation that preserves privacy.
		Therefore, we propose such a location privacy preference recommendation, referring to the method by Shin \cite{Shin2018PrivacyEM}.
		A comparison of our method with related work is shown in Table \ref{tab-placement}.
		
		\begin{table}[tb]
				\centering				
				\caption{A comparison of our method with related work \cite{zhao2014privacy}}\label{tab-placement}
				\begin{tabular}{|c|c|c|c|} \hline
					 & Method & How to preserve privacy \\ \hline
					 Related work \cite{zhao2014privacy} & 
					\begin{tabular}{c}
					Collaborative filtering\\(inaccurate for a large amount of data)
					\end{tabular}										 
					 & 
					 \begin{tabular}{c}
					 Add noise that is\\not strict mathematically 
					 \end{tabular}
					 \\ \hline
					 Our method & 
					\begin{tabular}{c}
					Matrix Factorization\\(accurate for a large amount of data)
					\end{tabular}						 
					&
					\begin{tabular}{c}
					Add noise based on\\local differential privacy
					\end{tabular}
					 \\ \hline
				\end{tabular}
			\end{table}
	
	\subsection{Location privacy preference definition}
	
		In location privacy preservation, it is important to define the privacy goal, that is, which information we protect.
		Such definition influences the recommendation of privacy preferences and the preservation of location privacy.
		There are ways to define a level that preserves location privacy regardless of location and time \cite{Andrs2013GeoindistinguishabilityDP} \cite{abul2008never} \cite{xiao_loclok:_2017}.
		The idea is to preserve location privacy at the same level at all locations and times.
		However, the users want to change the level according to their location and time.
		Therefore, we should define where and when we want to preserve location privacy, but there is no method for this definition.
		
		Cao et al. \cite{cao2018priste} \cite{cao_differentially_2015} \cite{cao_differentially_2016} proposed a method to define the privacy goal, which consisted of spatial and temporal goals.
		However, even in these studies, it was not mentioned about how to define information such as the place and time.
		Therefore, we propose a way to define such information in the best possible way.
		
\section{Preliminaries}\label{preliminaries}
	
	\subsection{Matrix factorization}\label{subsec-MF}
	
	    Matrix factorization is one of the most popular methods used for item recommendation, which predicts the ratings of unknown items.
		This is an extension of collaborative filtering to improve the accuracy for the large amount of data by dimentionality reduction.
		
		We consider the situation in which $m$ users rate any item in $n$ items (e.g., movies).
		We express each user's rating of each item by $\mathcal{M} \subset \{$$1 , \cdots, m \} \times \{ 1, \cdots, n \}$, 
		and the number of ratings as $M = | \mathcal{M} |$, for the user $i$'s rating of item $j$.
		Matrix factorization predicts the ratings of unknown items given $\{r_{ij} : (i, j) \in \mathcal{M}\}$.
		To make a prediction, we consider a ratings matrix $R = m \times n$, a user matrix $U = d \times m$, and an item matrix $V = d \times n$.
		The matrices satisfy the formula:
		$$
		R \approx U^T V
		$$
		
		The user $i$'s rating of item $j$ is obtained from the inner product of $U^T_i$ and $V_j$.
		In matrix factorization, a user element is expressed by $u_i \in R^d$, $1 \leq i \leq m$, and an item element is expressed by $v_j \in R^d, 1 \leq j \leq n$, which are learned from known ratings.
		In learning, we obtain the matrices $U$ and $V$, which minimize the following:
		\begin{equation}\label{eq-1}
		\frac{1}{M} \sum_{(i,j) \in \mathcal{M}} (r_{ij} - u^T_i v_j)^2 + \lambda_u \sum^m_{i=1} ||u_i||^2 + \lambda_v \sum^n_{j=1} ||v_j||^2
		\end{equation}
		$\lambda_u$ and $\lambda_v$ are positive variables for regularization.
		
		$U$ and $V$ are obtained by updating using the following formulae:
		\begin{equation}
		u^t_i = u^{t-1}_i - \gamma_t \cdot \{ \nabla_{u_i} \phi(U^{t-1}, V^{t-1}) + 2 \lambda_uU^{t-1}_i \}
		\end{equation}
		\begin{equation}
		v^t_j = v^{t-1}_j - \gamma_t \cdot \{ \nabla_{v_j} \phi(U^{t-1}, V^{t-1}) + 2 \lambda_vV^{t-1}_j \}
		\end{equation}
		$\gamma_t$ is the learning rate at the $t$th iteration, and $\nabla_{u_i} \phi(U, V)$ and $\nabla_{v_j} \phi(U, V)$ are the gradients of $u_i$ and $v_j$.
		They are obtained from derivative of (1) and expressed by the followings:
		\begin{equation}
		\nabla_{u_i} \phi(U, V) = - \frac{2}{M} \sum_{j:(i,j) \in \mathcal{M}} v_j(r_{ij} - u^T_i v_j)
		\end{equation}
		\begin{equation}\label{eq-nablav}
		\nabla_{v_j} \phi(U, V) = - \frac{2}{M} \sum_{i:(i,j) \in \mathcal{M}} u_i(r_{ij} - u^T_i v_j)
		\end{equation}
		
		We predict the ratings of the unknown items by calculating $U$ and $V$ by these formulae.

	\subsection{Local differential privacy}
	    We use local differential privacy to expand the matrix factorization into a form that satisfies privacy preservation.
		This approach is an extension of differential privacy \cite{dwork2008differential},
		which is a mechanism that protects the sensitive data in databases from attackers but enables accurate statistical analysis of the entire database \cite{nakagawa2016privacy}.
		In differential privacy, a trusted server adds noise to the data collected from the users.
		However, we assume that the server can not be trusted and use local differential privacy, in which users add noise to the data before sending the data to the server.
		
		The idea behind realizing local differential privacy is that for a certain user, regardless of whether or not the user has certain data, the statistical result should not change.
		The definition is given below:
		\newtheorem{defi}{Definition}
		\begin{defi}[Local differential privacy]
		We take $x \in N$，$x' \in N$.
		A mechanism $M$ satisfies $\epsilon$-local differential privacy if $M$ satisfies the following:
		$$
		Pr[M(x) \in S] \le \exp(\epsilon)Pr[M(x') \in S]
		$$
		\end{defi}
		$\forall S \subseteq Range(M)$ is any output that $ M $ may generate.
		A randomized response \cite{warner1965randomized} is used to realize local differential privacy, which decides the value to output based on the specified probability when inputting a certain value.
		Each user can add noise to the data according to their own privacy awareness, since he/she can decide the probability.
	
	\subsection{Definition of the location privacy preference}
		
		The location privacy preference is defined by the following:
		\begin{defi}[Location privacy preference]
		    The location privacy preference $p_u(t,p)$, in which the user $u$ wants to hide location information at time $t$ in location $l$, is expressed by the following:
			\begin{eqnarray}
				p_u(t,l)=\left\{ \begin{array}{ll}
				1 \, (Positive)\\
				0 \, (Negative)\\
				\end{array} \right.
			\end{eqnarray}	
			$1 \, (Positive)$ means that he/she can publish location information, and $0 \, (Negative)$ means that he/she does not want to publish location information.
		\end{defi}
		The time $t$ is expressed as a slot of time divided by a certain standard.
		The division method varies depending on the reference time, but as an example, it can be divided as follows:
		$
		\{ Morning, \, Noon, \, Afternoon, \, Evening, \, Night, \, Midnight \}
		$
		Additionally, the location $ l $ is represented by a combination of geographic information and a category of place or either of these.
		Geographical information represents the latitude and longitude or certain fixed areas, and the category represents the property of a building located in that place such as a restaurant or a school.
		
        There are various granularity regarding how to represent this information.
		As for the time $t$, ``February 1st, 2019" is expressed by ``Afternoon",  ``Weekday, Afternoon", and ``February, Weekday, Afternoon".
		As for the location $l$, ``Charleston Airport (Address : 5500 International Blvd, Charleston, SC 29418, the US)" is expressed by ``Airport" or ``Airport, Charleston".
		As the granularity of information changes, the number of goods in the recommendation and the degree of consideration of the nature of the commodity change, which influence the utility of the recommendation.
		However, it is not clear what kind of granularity is the best.
		Therefore, we confirm the best granularity, that is, the definition of best location information privacy preferences.
		
\section{Recommendation method}\label{sec-method}
	\subsection{Framework}\label{subsec-overview}
	
		We propose a location privacy preference recommendation method that preserves privacy.
		When a user enters location privacy preferences for a certain number of places and time combinations, the method outputs location privacy preferences for a combination of unknown places and times.
		We assume the recommendation system exists on an untrusted server and an attacker who tries to extract the location and time of the user visit and its rating based on the output of the system.
		Our method aims for compatibility between high availability as a recommendation scheme and privacy protection.
		We realize the former by matrix factorization and the latter by local differential privacy.
		
		First, we show a rough flow for the recommendation of the location privacy preference using the normal matrix factorization below.
		
		\begin{enumerate}
			\item The recommendation system sends the user matrix $U$ and item matrix $V$ to the user.
			\item The gradients are calculated by using the user's data and step 1 and sent to the system.
			\item The system updates $U$ and $V$ by the gradients calculated in step 2.
		\end{enumerate}
		
		This operation is performed a number of times, and there is a risk that the user's data is leaked to the attacker.
		
\begin{figure}[tb]
		\begin{center}
		\includegraphics[width=\textwidth]{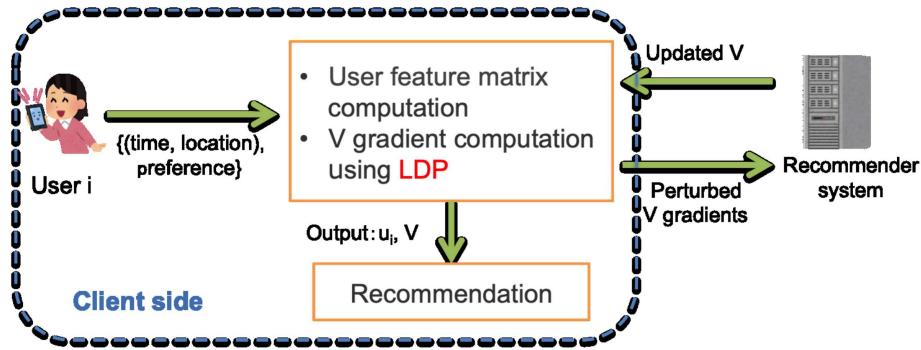}
		\caption{Overview of our method}\label{fig-overview}
		\end{center}
		\end{figure}

		Therefore, we add noise to the data in step 2 above to avoid such risks.
		When the user calculate the gradients to update $ U $ and $ V $, noise that satisfies local differential privacy is added to the information regarding ``when and where the user visits" and ``whether he/she wants to publish the location information".
		The gradients calculated based on the noise-added data are sent to the recommendation system.
		We show the overview of our method in Fig. \ref{fig-overview}.
		
		If we repeat the above steps $ k $ times, the gradient should be different each time.
		Therefore, there is a risk that some information can be obtained by the attacker by comparing each gradient.
		To avoid such an attack, we propose a method that satisfies $\epsilon$-local differential privacy for all combination of the gradients.
	
	\subsection{Noise adding}\label{subsec-detail}

		We preserve privacy for the information, that is, when and where the user visits and whether he/she wants to publish their location information.
		In privacy protection process, we refer the method by Shin \cite{Shin2018PrivacyEM}.
		
		First, we will describe how to add noise to the information regarding the time and location of the user's visits.
		Let $ y_{ij} $ be a value indicating whether or not user $ i $ has visited place $ j $, which is $1$ if he/she has visited the place and $0$ otherwise.
		The following equation holds:
		$$
		\sum_{(i,j) \in \mathcal{M}} (r_{il} - u^T_i v_j)^2 = 
		\sum^n_{i=1} \sum^m_{j=1} y_{ij}(r_{ij} - u^T_i v_j)^2
		$$
		Therefore, equation (\ref{eq-nablav}) can be transformed as follows.
		\begin{equation}
			\nabla_{v_j} \phi(U, V) = - \frac{2}{n} \sum_{i:(i,j) \in \mathcal{M}} y_{ij} u_i(r_{ij} - u^T_i v_j)
		\end{equation}
		To protect information regarding the time and location of the user's visits from privacy attacks, we should add noise to a vector $Y_i = (y_{ij})_{1 \leq j \leq m}$ representing whether or not the user has evaluated the item.
		We use a randomized response, and the value $y^*_{ij}$ obtained by adding noise to $y_{ij}$ is obtained as follows.
		\begin{eqnarray}
			p_u(t,p)=\left\{ \begin{array}{ll}
			0, \, with \, probability \, p/2\\
			1, \, with \, probability \, p/2\\
			y_{ij}, \, with \, probability \, 1-p
			\end{array} \right.
		\end{eqnarray}
	
	    Next, we describe a method of privacy protection for information on whether to disclose location information for a certain place and time combination.
		We add noise $\eta_{ijl}$ to the value $g_{ij} = (g_{ijl})_{1 \leq l \leq d} = -2 u_i(r_{ij} - u^T_i v_j)$.
		The noise is based on a Laplace distribution, and the noise-added value $g^*_{ijl}$ is expressed as follows:
		\begin{equation}
			g^*_{ijl} = g_{ijl} + \eta_{ijl}
		\end{equation}
		
		Each user adds noise to his/her own data in the above way, and the noise-added gradients, $\{ (y^*_{ij} g^*_{ij1}, \dots g^*_{ijd}) : j = 1, \dots , m \}$, are sent.
		The gradient used for updating the matrix $ V $ is the average value of the gradient of each user, expressed by $\nabla_{v_j} = n^{-1} \sum^n_{i=1} y_{ij} g_{ij}$.
		Then, the noise added to this value is expressed as follows.
		\begin{equation}
			\nabla^*_{v_j} = \frac{1}{n} \sum^n_{i=1} (\frac{y^*_{ij} - p/2}{1-p}) g^*_{ij}
		\end{equation}
		
		By repeating the operation $ k $ times, updating the value of the matrix using the slope calculated using the data with noise added, we find the matrix for predicting the evaluation value.
			
\section{Evaluation}\label{sec-eval}
	\subsection{Overview}\label{subsec-evaloverview}
	
	    In this section, we describe the evaluation indices and points of view to consider when verifying the utilities of our method.
		
		We evaluate the following two points, the approximation between the true ratings value and the predicted one, and accuracy of the prediction.
		We describe the details of these metrics in \ref{subsec-metrics}.
		
		In the evaluations, we compare the utilities of the recommendation methods using normal matrix factorization and local differential privacy.
		In addition, we evaluate the method from the following three viewpoints:
		
		\begin{enumerate}
			\item What is the best location privacy preference definition?
			\item How much impact does changes in the privacy preservation level make?
			\item What impact does changes in the number of unknown evaluation values make?
		\end{enumerate}
		
	\subsection{Dataset}\label{subsec-eval-processing}
	
		In the evaluation, we use artificial data combining the location privacy preference dataset and the position information dataset.
		
		For the location privacy preference dataset, we use LocShare acquired from the data archive CRAWDAD \cite{locShare}.
		This dataset was obtained from 20 users in London and St. Andrews over one week from April 23 to 29in 2011, with privacy preference data for 413 places.
		This dataset has few users, so it is not suitable for the evaluation of our method using matrix factorization \cite{koren2009matrix}.
		In addition, we need bulk data in the evaluation because there are many users of recommendation in the real world.

		Therefore, we generated an artificial dataset by combining the location privacy preference dataset with the trajectory dataset Gowalla, which was acquired from the location information SNS in the US.
		This dataset includes check-in histories of various places from 319,063 users collected from November 2010 to June 2011.
		The total number of check-ins is 36,001,959, and the number of checked-in places is 2,844,145.

	\subsection{Metrics}\label{subsec-metrics}
		
		We describe the metrics for verifying the utility of our method.
		
		\noindent
		{\bf The approximation when comparing the entire matrix:}
			We measure the similarity of the two matrices, the matrix $ R $ representing the true evaluation value and the matrix $ U ^ T V $ representing the estimated one, as a whole.
			
			Originally, the evaluation value is expressed by either $ 1 $ (Positive: can be released) or $ 0 $ (Negative: cannot be released).
			However, the product of the matrix $ U ^ T $ and $ V $ is not one of these but is classified as either $ 1 $ or $ 0 $ with a certain threshold, which is the mean of the product elements.
			We measure how accurately the recommendation can predict the ratings.
			The predicted value in the recommendation can be classified based on the true evaluation value as in Table \ref{tab-true-positive}.
			For example, TP (True Positive) is the number of data that are truly Positive (can be released) and whose predicted values are also Positive.
			We calculate the number of TP, FP, TN, and FN results from the prediction result and measure the following two indices.
			
			\begin{table}[t]
				\centering				
				\caption{A measure representing true or false values of the result}\label{tab-true-positive}
				\begin{tabular}{|c|c|c|c|} \hline
					\multicolumn{2}{|c|}{} & \multicolumn{2}{|c|}{True rating} \\ \cline{3-4}
					\multicolumn{2}{|c|}{} & Positive & Negative \\ \hline
					Predicted rating & Positive & TP (True Positive) & FP (False Positive) \\ \cline{2-4}
					 & Negative & FN (False Negative) & TN (True Negative) \\ \hline
				\end{tabular}
			\end{table}
			
			\begin{itemize}
			\item False Positive Rate: 
				The false positive rate is the percentage of false positives predicted relative to the number of negatives.
				$$
				FPR = \frac{FP}{TN + FP}
				$$
				This metric is an index for verifying whether the location information that the user wants to disclose is not erroneously disclosed.
				The lower the false positive rate, the higher the accuracy of the privacy protection.
			
			\item Recall: 
				Recall is the proportion of data predicted to be positive out of the data that are actually positive.
				$$
				Recall = \frac{TP}{TP + FN}
				$$
				Recall is an index for verifying whether the location information that the user can publish is predicted to be positive, since if the released location information decreases, the benefit decreases.
				The higher the recall value is, the higher the utility of the recommendation.
		    \end{itemize}
		
\begin{figure}[t]
			\begin{center}
			\includegraphics[width=12cm]{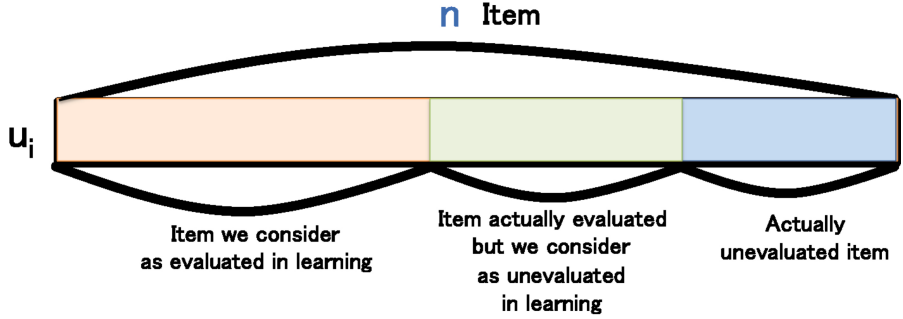}
			\caption{An example in which a part of the evaluation value of some users is regarded as unevaluated. In the evaluation, we measure how correctly the yellow-green part is predicted.
			}\label{fig-ukrate}
			\end{center}
		    \end{figure}

		\noindent
		{\bf Accuracy of the prediction for each user:}
            In this experiment, as shown in Fig. \ref{fig-ukrate}, a part of the products considered unevaluated is taken as learning data for some users.
			We pay attention to the result of an item that is regarded as unevaluated in recommendation.
			In such cases, we measure the ratio of correctly predicted evaluations relative to the true evaluation values.
			We call this ratio the reconstruction rate, which can be calculated by the following:

			$$
			Reconstruction \, rate = \frac{TP + TN}{TP + FP + TN + FN}
			$$

            The higher the reconstruction value, the more useful the recommendation.
			
	\subsection{Evaluation process}\label{subsec-evalmethod}
	
	    We evaluate our method from the three viewpoints mentioned in \ref{subsec-evaloverview}, and we adopt each of the following methods.
		
		\begin{enumerate}
			\item We used 10-fold cross validation, that divides the users into training data and test data , in which 
			90\% of the users are regarded as training data and 10\% of the users are regarded as test data.
			\item Among the user's data included in the test data, we regard the known evaluation value as unknown according to the values of the $ Unknown \, Rate $ mentioned later.
			\item We predict the evaluation value by using the test data and the training data which have undergone conversion processing and calculate the metrics.
			\item We repeat the above process 100 times and verify the average of the evaluation indices.
		\end{enumerate}
		
		In the evaluation, we change the value of one of the following parameters: $time$, $\epsilon$, $Unknown \, Rate$.
		$time$ is the length of the standard when dividing time into multiple slots.
		$\epsilon$ is privacy protection level when using local differential privacy.
		$Unknown \, Rate$ is the ratio of what is regarded as unknown. 
	
	\subsection{Results}
		
		We describe the results of experiments to confirm the utility of the recommendation method.
	
\begin{figure}[h]
             \begin{minipage}[b]{0.5\textwidth}
              \centering
              \includegraphics[keepaspectratio, scale=0.45]
              {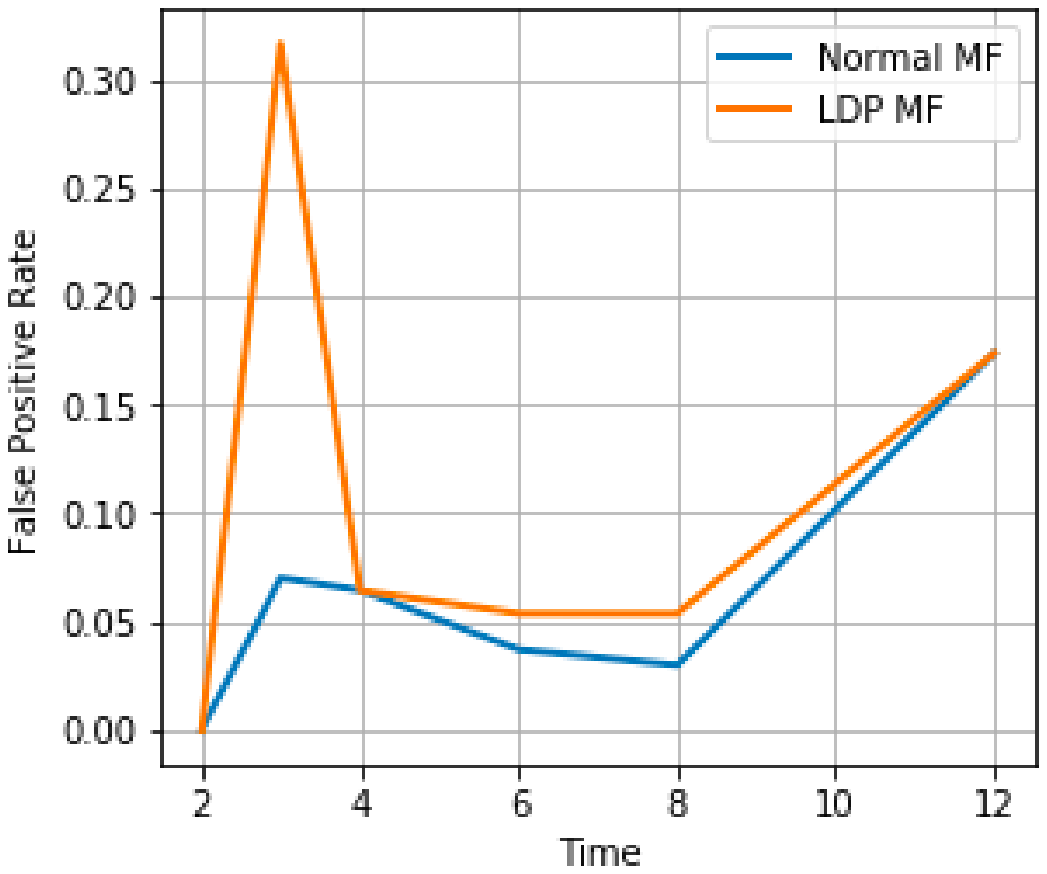}
              \subcaption{False positive rate}
             \end{minipage}
             \begin{minipage}[b]{0.5\textwidth}
              \centering
              \includegraphics[keepaspectratio, scale=0.45]
              {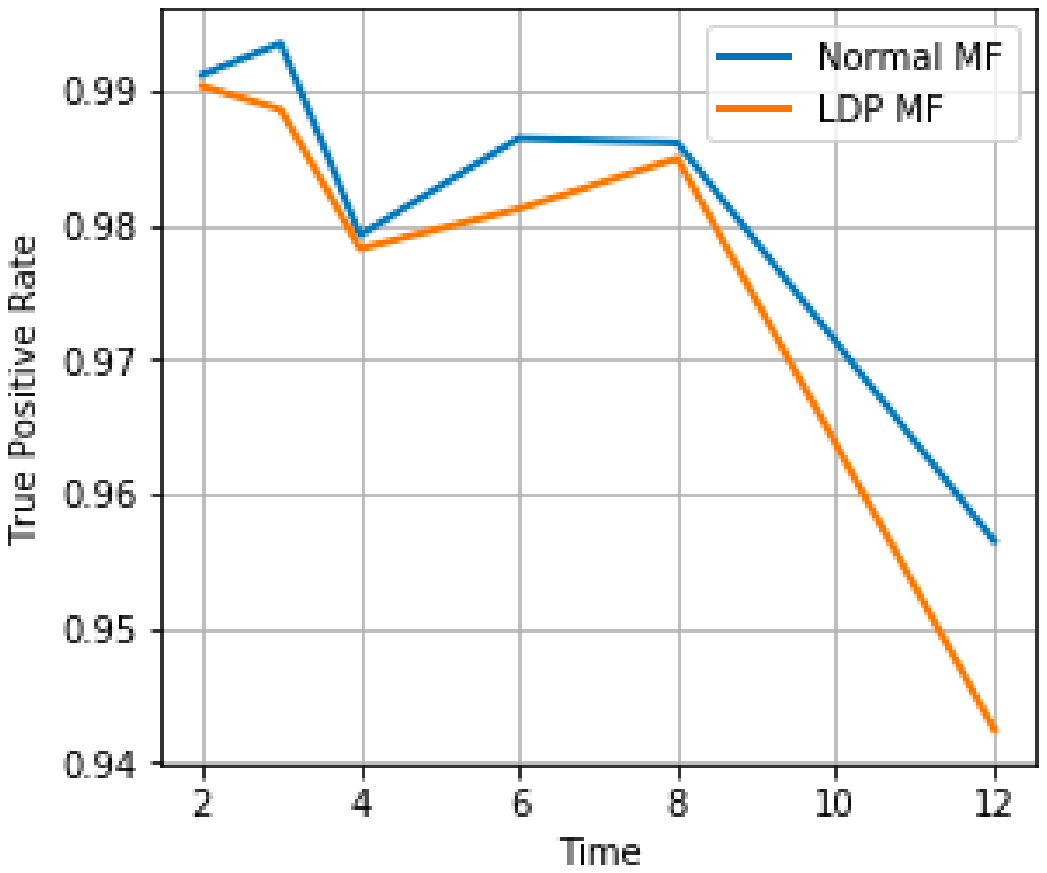}
              \subcaption{Recall}
             \end{minipage}\\
             \begin{minipage}[b]{0.5\textwidth}
              \centering
              \includegraphics[keepaspectratio, scale=0.45]
              {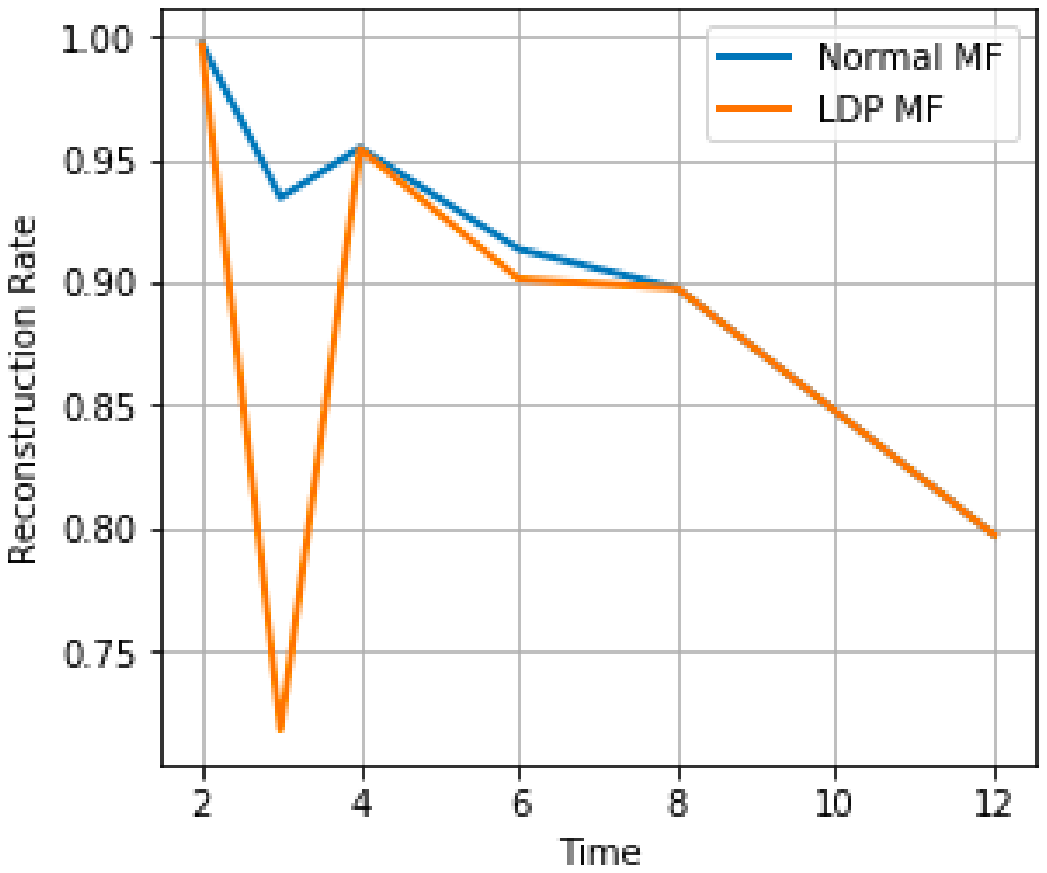}
              \subcaption{Reconstruction rate}
             \end{minipage}
             \caption{Results when the time granularity is changed.
             }\label{fig-tresult}
            \end{figure}

		\noindent
		{\bf The best location privacy preference definition:}
		    When defining the location privacy preference, the best definition regarding the granularity of the information is not yet clear.
			In this experiment, we examine the influence of changing the granularity of time on the utility.
			We change the criterion for dividing time into multiple slots as follows:
			$time = \, 2, \, 3, \, 4, \, 6, \, 8, \, 12,$
			the other parameters are set as follows: $\epsilon = 0.01$ and the $Unknown \, Rate = 0.1$.
			The results are shown in Fig. \ref{fig-tresult}.
			
			From these results, we confirm that the precision drops when the granularity is small, that is, when the criterion time is short.
			This is because as the granularity decreases, the number of goods in the recommendation increases, and the matrix used for the recommendation becomes sparse.
			On the other hand, however, we confirm that the accuracy drops even if the granularity is too large from the result of the output by the model of $time = \, 8, \, 12$.
			This is because the consideration of time becomes difficult.
			From these results, we confirm that we should control the granularity in the decision of the definition considering trade-offs of the utilities.
			
			When we compare the recommendations using normal matrix factorization and local differential privacy, there is not much difference in terms of utility.
			Therefore, the comparison also shows the possibility of an accurate recommendation while protecting privacy.
			
		\noindent
		{\bf Impact of changes in the privacy preservation level:}
		    The strength of the privacy protection can be adjusted with the value $ \epsilon $ in local differential privacy.
		    The smaller the value of $ \epsilon $, the more privacy is protected. On the other hand, there is a risk that the utility of the recommendation decreases as the added noise become large.
			Therefore, we examine the influence of changing the privacy protection level on the utility.
		
\begin{figure}[h]
             \begin{minipage}[b]{0.5\textwidth}
              \centering
              \includegraphics[keepaspectratio, scale=0.45]
              {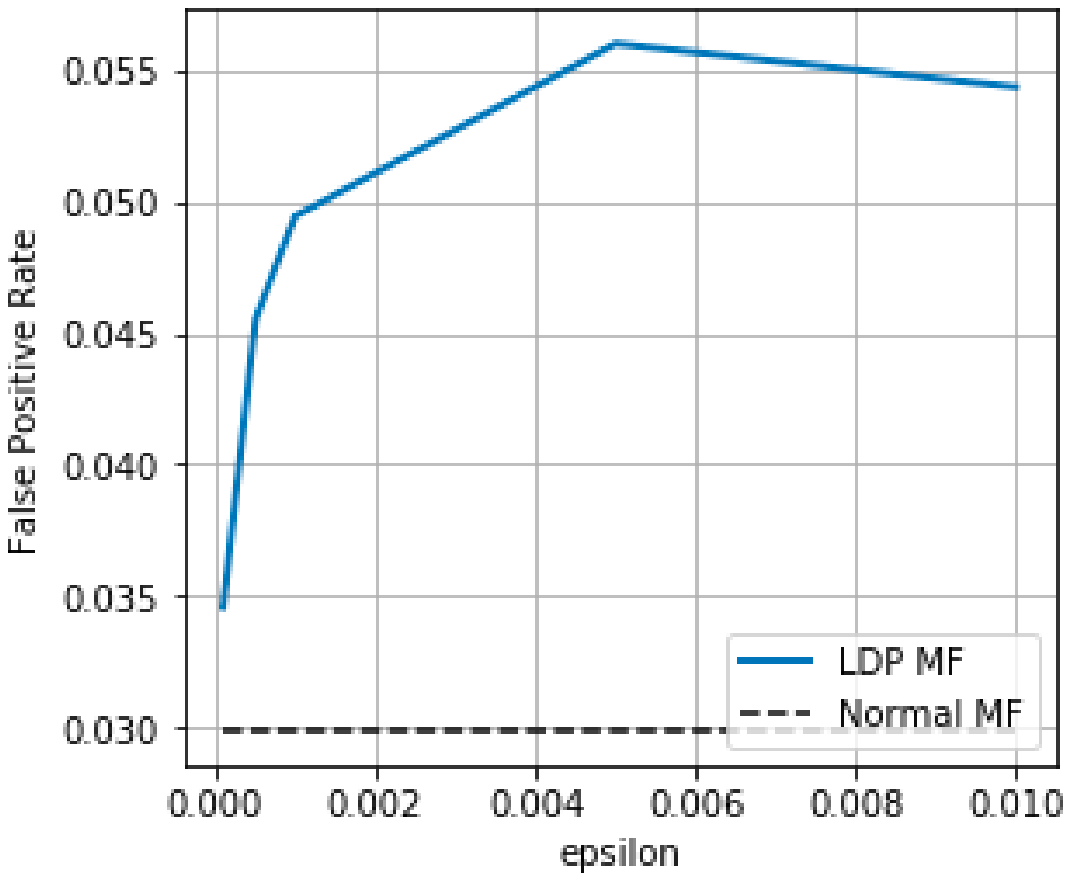}
              \subcaption{False positive rate}
             \end{minipage}
             \begin{minipage}[b]{0.5\textwidth}
              \centering
              \includegraphics[keepaspectratio, scale=0.45]
              {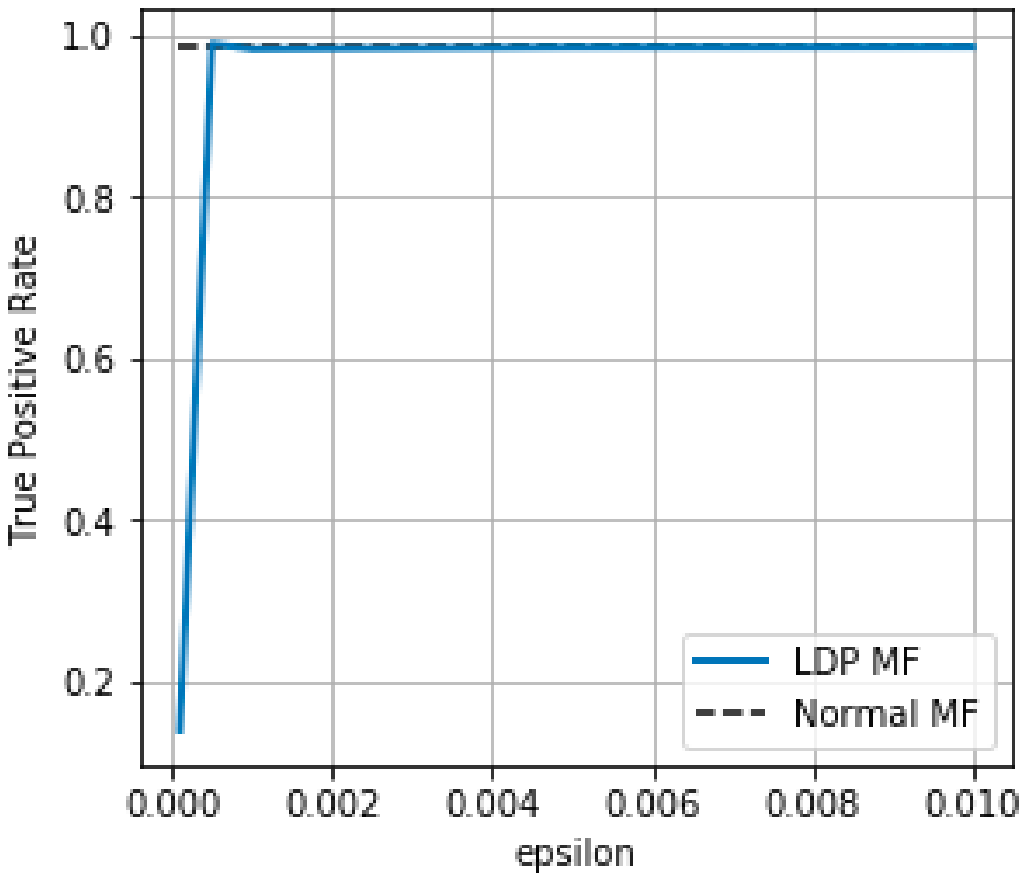}
              \subcaption{Recall}
             \end{minipage}\\
             \begin{minipage}[b]{0.5\textwidth}
              \centering
              \includegraphics[keepaspectratio, scale=0.45]
              {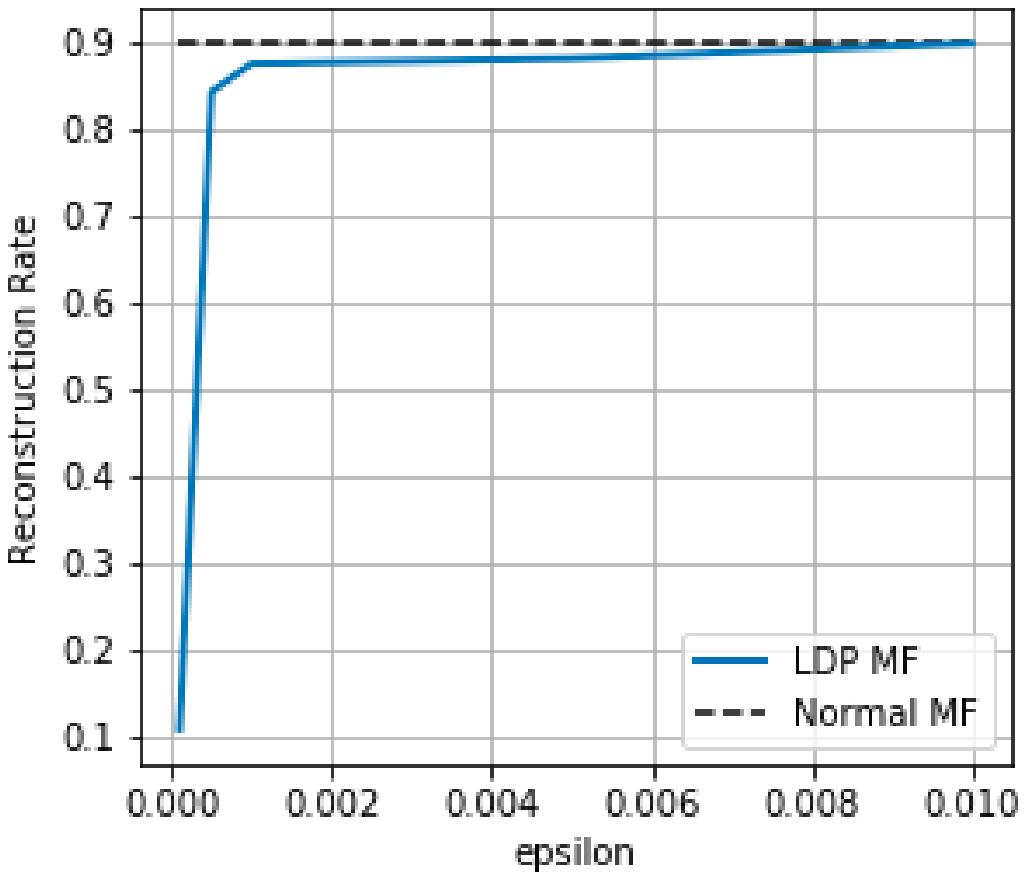}
              \subcaption{Reconstruction rate}
             \end{minipage}
             \caption{Results when the $\epsilon$ is changed.
             }\label{fig-eresult}
            \end{figure}

			In this evaluation, we change the privacy protection level as follows:
			$\epsilon = \, 0.0001, \, 0.0003, \, 0.001, \, 0.005, \, 0.01$,
			the other parameters are set as follows: $time = 6$ and the $Unknown \, Rate = 0.1$.
			The results are shown in Fig. \ref{fig-eresult}.

            From the results, we confirm that a normal recommendation is more useful in general, and as the value of $ \epsilon $ increeases, the usefulness increases.
			On the other hand, however, the change in the usefulness due to the change in the value of $ \epsilon $ is small for $ \epsilon = 0.003 $.
			Larger values do not have a significant effect on the usefulness.
			We should set the value such that the effect of the decrease in the usefulness becomes small, since the privacy protection level is higher when $\epsilon$ is small.
			
\begin{figure}[h]
             \begin{minipage}[b]{0.5\textwidth}
              \centering
              \includegraphics[keepaspectratio, scale=0.45]
              {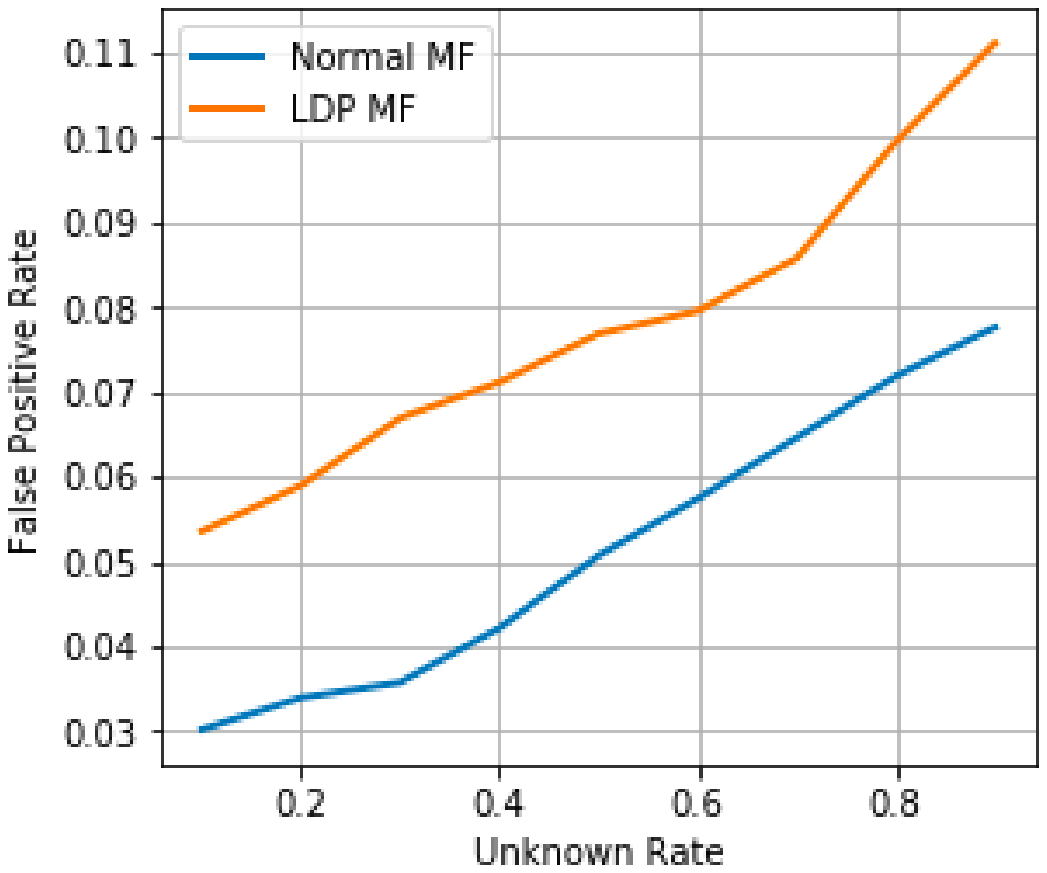}
              \subcaption{False positive rate}
             \end{minipage}%
             \begin{minipage}[b]{0.5\textwidth}
              \centering
              \includegraphics[keepaspectratio, scale=0.45]
              {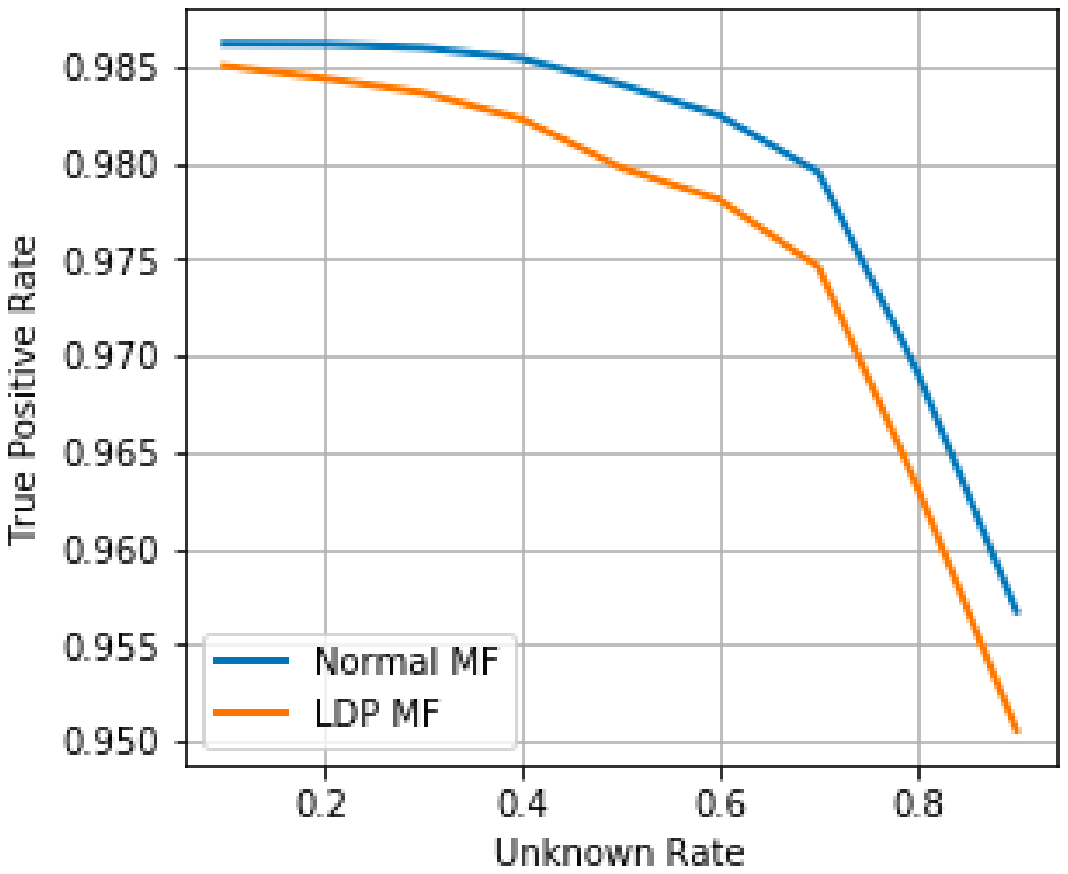}
              \subcaption{Recall}
             \end{minipage}\\
             \begin{minipage}[b]{0.5\textwidth}
              \centering
              \includegraphics[keepaspectratio, scale=0.45]
              {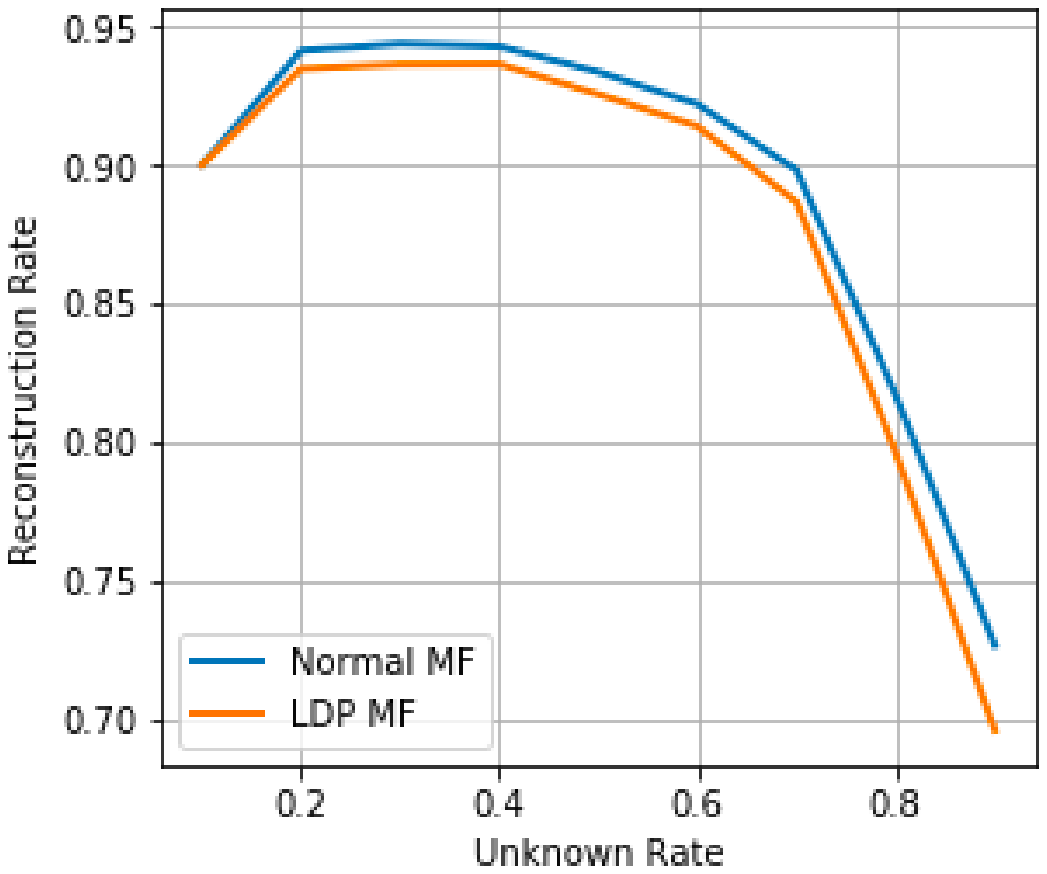}
              \subcaption{Reconstruction rate}
             \end{minipage}
             \caption{Results when the $Unknown \, Rate$ is changed.
             }\label{fig-nresult}
            \end{figure}

		\noindent
		{\bf Impact of changes in the number of unknown evaluation values:}
		    We verify how much each user should know his/her privacy preference for an accurate prediction.
		    In the evaluation, we regard a certain number of evaluated data as unevaluated in generating the model and verify the influence of the number of unknown evaluation values on the utility.
			We change the parameter for the percentage of unevaluated data as follows:
			$Unknown \, Rate = 0.1, \, 0.2, \, 0.3, \, 0.4, \, 0.5, \, 0.6,\\ \, 0.7, \, 0.8, \, 0.9$
			The smaller the $Unknown \, Rate$, the higher the number of evaluation values is.
			The other parameters are set as follows:
			$time = 6$ and $\epsilon = 0.01$.
			The results are shown in Figs. \ref{fig-nresult}.
			
			From these results, we confirm that the utility tends to decrease as the number of unevaluated data values increases.
			This is because the accuracy of the recommendation will be reduced if the training dataset is small.
			On the other hand, from the result of reconstruction rate, when more data are learned, it will be more likely that it will be difficult to make accurate recommendations.
			This is because overlearning occurs due to the large amount of learning data, and a model specialized for specific data is generated.
			
			From these results, a large amount of learning data is necessary in order to make accurate recommendations, but we have to control the amount of data to generate a versatile model.
			We also confirm that a user who does not have a large amount of data can use the recommendation accurately.
			
	\subsection{Discussion}
	
        In this section, we discuss the experimental results.
		
		First, we compare the utility of the models using normal matrix factorization and local differential privacy.
		Since appreciable differences were not observed in devising the parameters, we confirm that the recommendation method can maintain an accuracy comparable to that of normal matrix factorization.
		
		Next, we discuss about the influence of changing the following parameters: the granularity of the location privacy preference, the privacy protection level, and the number of unknown ratings.
		The best parameter in this evaluation can change according to the data, but the method of selecting the parameter can be applied to any data.
		
		As for the granularity of the location privacy preference, as the granularity becomes coarse, that is, the criterion for dividing time into multiple slot becomes long, the utility improves and becomes worse at a certain value.
		This is because the coarser the definition of the granularity is, the smaller the number of goods in the recommendation, and the matrix used for prediction becomes dense.
		On the other hand, however, we confirm that when the granularity is too coarse, it becomes impossible to properly consider the property of time, and the accuracy decreases.
		Therefore, in defining the location privacy preference, we should choose the criterion with the highest utility.
		In this evaluation, the best criterion is 8 hours.
		
		Regarding the privacy protection level, the utility improves as the value of $ \epsilon $ increases, but the increasing rate of the utility increases from a certain value.
		We should select the maximum parameter that can maintain prediction accuracy, since a stronger privacy protection level is achieved for a smaller value of $ \epsilon $.
		Therefore, in this evaluation, the best privacy protection level is $\epsilon=0.001$.
		
		Finally, regarding the number of unknown ratings, we can determine how many ratings a new user of the recommendation system needs.
		Generally, we can make accurate precision for the large amount of training data.
		However, there is a tendency whereby the accuracy decreases when the number of rated items is too large.
		This is because overlearning occurs as a result of too much data.
		From these results, ideally, the user should know the location privacy preference nearly as much as possible, for about 70\% of all products.
		
		From the above discussion, we confirm that our method is compatible with usefulness and privacy protection and how to select each parameter.

\section{Conclusion}
	
	We propose a location privacy preference recommendation system that uses matrix factorization and achieves privacy protection by local differential privacy.
	We also confirm how to determine the best location privacy preference definition.
	
	In the evaluation using an artificial dataset from a location privacy preference dataset and a trajectory dataset, we evaluate our method by using three metrics: a false positive rate, recall, and a reconstruction rate.
	From the results of the evaluation, we confirm that our method can maintain the utility at a level that is the same as a method without privacy preservation.
	In addition, we confirm that the best parameters are the granularity of the location privacy preference definition, the privacy protection level, and the number of rated items.

	Our future works are listed below:
	
	\begin{itemize}
		\item The best location information granularity in the location privacy preference definition: 
		We examined the definition of time in the location privacy preference.
		However, it is also necessary to clarify the method of defining the location.
		We will confirm the effectiveness of two pieces of information: geographical information such as latitude and longitude and category information of the place.
	
		\item Extension of the recommendation method considering features of the location information: 
		Although we confirmed that the utility was inferior to that of normal matrix factorization, there is room for improvement.
		One idea to be considered is extending the consideration of the features of the position information when we use matrix factorization.
		The idea is that when a user decides a location privacy preference, if there is one place that he/she wishes to hide, this place has an influence on nearby places.
		Actually, a study considered the influence of neighboring places in a system that recommends places to visit \cite{lian2014geomf}, although this was not research related to location privacy preferences.
		
		\item Application to a location privacy protection system: 
		If we use the location privacy preferences predicted by our method, a location privacy preserving system can protect location information without users' input.
		\item Dealing with data correlations: in the current method, we do not take the data correlations into account; however, previous studies \cite{cao_quantifying_2017} \cite{cao_quantifying_2018} show that differential privacy suffers from temporal privacy leakage. We will consider to use techniques  provided in \cite{cao_contpl:_2018} to solve this issue.
		
	\end{itemize}

%
%
%
\bibliographystyle{splncs04}
\bibliography{dbsec19-Asada}
%

\begin{subappendices}
	\renewcommand{\thesection}{\Alph{section}}%
	
\section*{A.  Data processing}

    In the evaluation, we generate an artificial dataset by combining a location information privacy preference dataset and a trajectory dataset.
	When joining the datasets, we use time information and category category information as a key, but the categories are different.
	The location privacy preference dataset has categories as follows:
	$$
	\{ Food \, and \, Drink, \, Leisue, \, Retail, \, Residential, \, Academic, \, Library \}
	$$
	The trajectory dataset has categories as follows:
	$$
	\{ Community, \, Entertainment, \, Food, \, Nightlife, \, Outdoors, \, Shopping, \, Travel \}
	$$
	Therefore, it is necessary to convert the data so that both datasets have the same category information.
	The specific generation method is as follows.
	
	\begin{enumerate}
		\item Preprocessing of the trajectory dataset\\
		We extract check-in data for places where two or more users visited from the trajectory dataset.
		
		\item Preprocessing of both datasets (conversion of the time information)\\
		We split the exact date and time into time slots with respect to the time information in both datasets.
		For example, when the data is divided every 6 hours, the slots can be divided as follows:
		$$
		\{ Morning, \, Noon, \, Afternoon, \, Evening, \, Night, \, Midnight \}
		$$
		In the evaluation, we divide the slots using multiple criteria: 2, 3, 4, 6, 8, and 12 hours.
		For a shorter criterion time, the recommendation obtains a larger number of goods.
		
		\item Preprocessing of both datasets (conversion of the location information)\\
		We converted the category information of the trajectory dataset, since we unify the categories in datasets with fewer categories as follows:
		\begin{itemize}
		\item $Community$\\
		We convert the information according to its lower category.
		A category is converted to ``$Residential$" if its lower category is ``$Home$".
		A category is converted to ``$Library$" if its lower category is ``$Library$".
		A category with a different lower category is converted to ``$Academic$".
		\item $Entertainment \, \to \, Leisure$
		\item $Food \, \to \, Food \, and \, Drink$
		\item $Nightlife \, \to \, Food \, and \, Drink$
		\item $Outdoors \, \to \, Leisure$
		\item $Shopping \, \to \, Retail$
		\item $Travel \, \to \, Leisure$
		\end{itemize}
		
		
		\item Generate the item data, combining location and time\\
		The items in the recommendation are represented by a combination of location and time.
		Therefore, we generate elements combining place and time for each dataset.
		
		\item Generate multiple user sets in the location privacy preference dataset\\
		We merge two datasets using the item data as a key.
		In the location privacy preference dataset, the total number of product data values possessed by each user is extremely small.
		Therefore, when users of both datasets are associated with one another in a one-to-one correspondence, the data to be generated become extremely sparse.
		To avoid this problem, we generate the set of multiple users of the location privacy preference dataset and associate one set with one user of the route dataset.
		The user set satisfies the condition that it has an evaluation value for all products when the users included in the set are combined.
		We generate a number of user sets satisfying this condition.
		
		\item Application of the location privacy preference dataset to the trajectory dataset
		We select one of the number of user sets generated in the previous stage randomly and merge each user in the route dataset using the item data as a key.
		
	\end{enumerate}
	

\end{subappendices}

\end{document}